\documentclass[epj]{svjour}
\usepackage{graphicx}

\begin{document}

\title{Power, Levy, Exponential and Gaussian Regimes
in Autocatalytic Financial Systems}
\author{Zhi-Feng Huang\inst{1}\and Sorin Solomon\inst{2}}
\institute{Institute for Theoretical Physics, Cologne University,
D-50923 K\"oln, Germany
\and Racah Institute of Physics, The Hebrew University, 
Jerusalem 91904, Israel}
\date{}
\abstract{
We study by theoretical analysis and by direct numerical
simulation the dynamics of a wide
class of asynchronous stochastic systems
composed of many autocatalytic degrees of freedom.
We describe the generic emergence of truncated power laws
in the size distribution of their individual elements.
The exponents $\alpha$ of these power laws are time independent
and depend only on the way the
elements with very small values are treated.
These truncated power laws determine the collective time evolution
of the system.
In particular the global stochastic fluctuations of the system
differ from the normal Gaussian noise according to the time and
size scales at which these fluctuations are considered.
We describe the ranges
in which these fluctuations are parameterized respectively
by: the Levy regime $\alpha < 2$, the power law decay with large 
exponent ($\alpha > 2$), and the exponential decay.
Finally we relate these results to the
large exponent power laws found
in the actual behavior of the stock markets and
to the exponential cut-off detected in certain recent
measurement.
\PACS{
{05.40.+j}{Fluctuation phenomena, random processes, noise, and 
Brownian motion}\and
{05.70.Ln}{Nonequilibrium thermodynamics, irreversible processes}
\and
{02.50.-r}{Probability theory, stochastic processes, and statistics}
     }
}
\titlerunning{Power, Levy, Exponential and Gaussian Regimes}
\authorrunning{Z.F. Huang and S. Solomon}
\maketitle
\section{Introduction}

It was realized since a very long time that the fluctuations 
$r(t)$ of stochastic systems made of many degrees of freedom 
are not generically distributed by Gaussian probability 
distributions \cite{Mandelbrot63,Stanley95}. 
On the theoretical side, P. Levy \cite{Levy37} discovered at 
the beginning of this century that the central limit theorem allows 
for a family of distributions which decay at infinity as a power
law: 
\begin{equation} 
P(r) \sim r^{-1-\alpha} \  \  {\rm with} \ \  0<\alpha <2 
\label{eq1} 
\end{equation} 
These distributions can be thought as the limit distributions for 
random walks with steps of sizes $s_i$ distributed by a power law 
$$
P(s) \sim s^{-1-\alpha}.
$$
Such processes were named Levy flights \cite{Schlesinger95}.

In nature, the situation turns out to be more complicated: 
many of the measurements produced probability distribution functions 
which look like Levy distributions for a certain range of the 
stochastic variable but are cut off. That is, they change their 
character above a certain threshold, e.g., by becoming exponentials 
or changing to a power decay with $\alpha>2$ as found for the returns of
stock market data \cite{Stanley95,Lux96,Gopikrishnan99,Huang00}. 
Particular examples of such distributions were studied under the 
name of truncated Levy flight \cite{Mantegna94}.
As required by the central limit theorem, for very large time 
intervals $t$, these distributions cross-over into a quite Gaussian 
behavior.

Recently, both the power law probability distribution of the 
individual steps and the truncated Levy distribution of the 
fluctuations were explained generically by a series of Generalized 
Lotka-Volterra (GLV) models \cite{Solomon96,Solomon98,Biham98,Malcai99}.
These models represent realistic financial, biological and social 
systems composed of many autocatalytic and competing stochastic 
subsystems \cite{Solomon98}. 

In this paper we study the exact nature of the cut-off region and
the details of the cross-over process in the framework of the GLV
models. We especially describe the emergence of
the tail distribution with  $\alpha \approx 3$
power law and the exponential cut-off. The theoretical analysis below 
yields well defined quantitative predictions on the various temporal 
and geometric properties of the probability distribution functions, 
which are verified by the numerical simulations and compared with the 
actual measurements of the Hong Kong stock market.
Most of the results extend however to other models too. 

\section{Theoretical analysis}

\subsection{Autocatalytic systems and power law}

The analysis here applies to a large range of 
dynamical models \cite {Solomon98}. 
For definiteness we consider a specific system 
$W(t)$ formed of subsystems $w_i (t)$, $i= 1,2,...,N$: 
\begin{equation} 
W (t) = w_1(t)+w_2(t)+...+w_N(t) 
\label{eq2} 
\end{equation} 
distributed by a power law cumulative distribution function: 
\begin{equation} 
P(> w) = [w/w_{min}(t)]^{-\alpha} 
\label{eq3} 
\end{equation} 
with the corresponding density $P(w) \propto w^{-1-\alpha}$,
where $w_{min}$ is some lower cut-off 
\begin{equation}
w_{min} (t) = q \bar{w} (t),
\label{eq4} 
\end{equation}
i.e., usually given in terms of a fraction $q$ of the average 
$\bar{w} (t)=W(t)/N$.

Such distributions emerge naturally in 
autocatalytic systems \cite{Solomon96,Solomon98,Biham98,Malcai99}
of the generic form:
\begin{equation}
 w_i  (t+1) = \lambda (t) w_i (t) + a {\bar w} (t) 
- b  w_i (t) {\bar w} (t).
\label{eqsorin}
\end{equation}
For the purpose of this paper we describe them 
by a simple dynamics \cite{Solomon96,Malcai99} which consists in 
choosing randomly at each time step for updating one of the 
$w_i$'s and multiplying it by a random factor $\lambda(t)$ extracted 
at each time from a probability distribution $\rho(\lambda)$: 
\begin{equation} 
w_i  (t+1) = \lambda (t) w_i (t),          
\label{eq5} 
\end{equation} 
with the lower cut-off 
\begin{equation} 
w_i  (t+1) \geq q \bar{w} (t).
\label{eq6} 
\end{equation}
That is, the updated variable $w_i(t+1)$ is constrained to be
not smaller than a lower bound $q\bar{w}(t)$, i.e.,
$$
w_i(t+1)={\rm max}\{w_i(t+1),q\bar{w}(t)\},
$$
where $\bar{w}(t)$ is the average value calculated at earlier time $t$.
Note that all the $w_i$'s 
are strictly positive and therefore
\begin{equation} 
w_i (t) < W(t) = N  \bar{w} (t).
\label{eq7} 
\end{equation}

In systems of the type Eqs. (\ref{eq5}) and (\ref{eq6}), it has been 
shown \cite{Solomon96,Malcai99} that even if the system is not 
stationary the power law Eq. (\ref{eq3}) holds, and for given $N$ 
and $q$ in the range $1>q\gg 1/\ln N$, the exponent $\alpha$ is given 
by the relation: 
\begin{equation} 
\alpha = 1/(1-q) 
\label{eq8} 
\end{equation}
Usually in financial applications $q\sim 1/3$, and then the exponent
$\alpha \sim 1.5$ within the stable Levy regime.

During our analysis of this section we will take the distribution 
$\rho(\lambda)$ centered around 1: 
\begin{equation} 
\lambda(t) = 1+g(t),
\label{eq9} 
\end{equation} 
and consisting of just 2 equally probable values 
\begin{equation} 
g(t) = \pm M.
\label{eq10} 
\end{equation}
The conclusions of our analysis are not changed if 
one assumes an arbitrary distribution $\rho(\lambda)$ 
with finite strictly positive support.

Here we are interested in the distribution of the "returns" : 
\begin{equation} 
r (\tau) = [W(t+\tau) - W(t)]/W(t)        
\label{eq11} 
\end{equation} 
as a function of the time interval $\tau$. The use of the term 
"return" in (\ref{eq11}) as a measure of the fluctuation in the 
system is borrowed from the financial applications where $W(t)$
is the stock market index and $r(\tau)$ is the relative gain/loss
one incurs after a time lapse $\tau$.

Since the variation of $W(t)$ at each time coincides with 
the variation of the individual $w_i (t)$ which happened to be 
randomly selected for updating by Eq. (\ref{eq5}) at time $t$, 
the value of $r(\tau)$ as defined in Eq. (\ref{eq11}) 
is the result of a random walk 
\begin{equation} 
r (\tau) = \sum\limits_{k} s_i (t+k),
\label{eq12}
\end{equation}
with steps $s_i (t+k)$, $k=0,1, ...,\tau-1$ of sizes \cite{comment1}: 
\begin{equation} 
s_i (t+k) = [w_i (t+k+1) - w_i (t+k)] / W(t),
\label{eq13} 
\end{equation} 
which according to Eqs. (\ref{eq5}), (\ref{eq9}) and  (\ref{eq10})
is written as: 
\begin{equation} 
s_i (t+k) =  \pm M w_i (t+k)/W(t).
\label{eq14} 
\end{equation} 
Therefore, the sizes of the (absolute values of the) individual steps 
$s_i (t+k)$ in the random walk/flight process $r(\tau)$ (Eq. 
(\ref{eq12})) have a probability distribution function similar to 
Eq. (\ref{eq3}) (up to the factor $M/W$): 
\begin{equation} 
S(> s) = [s/s_{min}]^{-\alpha},
\label{eq15} 
\end{equation} 
where $s_{min} = M w_{min} / W = M q /N$.

\subsection{Origin of truncation in autocatalytic systems}
\label{sec2.2}

A crucial fact of the present paper is that the power law 
distribution Eq. (\ref{eq15}) from which the individual steps 
Eq. (\ref{eq13}) composing the random walk $r (\tau)$ through 
Eq. (\ref{eq12}) are selected is {\it truncated from above}
\cite{note2}. 

Indeed, since $w_i  < W$ for any $i$ as shown in Eq. (\ref{eq7}), 
the individual steps Eq. (\ref{eq13}) cannot be larger than the 
fixed value $M$, i.e. 
\begin{equation} 
S(>M) = 0.
\label{eq16} 
\end{equation} 
Note that this bound in the size of the individual steps 
of the random walk $r(\tau)$ does not depend on the number 
$N$ of participants in the game nor on the exponent 
$\alpha$ of the power law, nor on the lower cut-off $w_{min}$. 
One cannot therefore hope that the effects of this 
upper bound would somewhat become irrelevant, 
and as seen below (Eq. (\ref{eq25})), by increasing 
$N$ one can only delay the time 
$$
\tau_{max} = N^\alpha
$$
by which these effects become dominant.
Therefore, for finite time intervals $\tau$ and infinite $N$, the 
effect of the truncation disappears (together with the vanishing of 
the amplitude of  elementary fluctuations $|r(1)| < N^{1-\alpha}$ 
\cite{Solomon98}).
In finance, however, we are generically in the opposite limit,
e.g. a finite number of traders trading for very long time intervals 
$\tau$ which allow them to perform a total number of elementary 
trading operations much larger than $\tau_{max}$.

Our goal in the sequel will be therefore to follow in detail the 
process by which the $\tau=1$ (truncated at $r>M$) power 
distribution 
\begin{equation} 
R(>r,1) = S(>r) = [r/s_{min}]^{-\alpha} = [r/(qM/N)]^{-\alpha} 
\label{eq17} 
\end{equation}
evolves for increasing time interval $\tau$, and then approaches the 
infinite time ($\tau \gg \tau_{max}$) Gaussian distribution. 
We will analyze explicitly $R(>r,\tau)$ only for positive $r$'s to 
avoid unnecessary complication. However, the analysis for negative 
$r$ values is very similar. 

In fact, for the probability distribution function of returns 
$R(>r,\tau)$ one obtains a symmetric {\it probability density} 
$R(r,\tau) = {d R(>r,\tau)} / {d r}$ which for $\tau=1$ 
coincides with the $S(s) ={d S(>s)}/{d s}$ probability density. 
On a log-log scale the probability density $\ln R(r,1)$ vs. 
$\ln r$ is a line ending sharply around $r=s_{min}$. 
We will see that for larger $\tau$ values the sharp tip 
will erode into a flatter "dome" and the complete/exact vanishing 
of the distribution at the upper cut-off $r=M$ will evolve into a 
steep but continuous decay. 

\subsection{Power law and truncation for $\tau < \tau_{max}$}
\label{sec2.3}

The probability $R(>r,\tau)$ to reach after a $\tau$-steps walk 
a distance $r$ or larger is of course a result of the probabilities 
of the $\tau$ individual steps which compose the 
"walking"/"traveling"/"flying" \cite{Mantegna94} process. 
Therefore its characteristics depend on the 3 crucial properties 
of the individual steps distribution $S(>s)$ obtained from 
Eqs. (\ref{eq15})--(\ref{eq17}):

\begin{enumerate} 

\item The great majority of the individual steps 
$s_i$ are of order $s_{mean}$ (the average of $s_i$) and less, 
i.e., in the range: 
\begin{equation} 
s_{min} = q s_{mean} = q M/N < s < 2 s_{mean} = 2M/N.     
\label{eq18} 
\end{equation}
 
\item  The steps of larger sizes (say larger than $2 s_{mean}$) 
are very rare: 
\begin{equation} 
R(>2M/N,1) \sim (q/2)^{\alpha} < 0.07
\label{eq19} 
\end{equation}
(We take $\alpha \sim 1.5$ as in real wealth distribution
\cite{Levy97}).
 
\item  There are {\it no} individual steps in $r(\tau)$ of size larger 
than $M$: 
\begin{equation} 
R(> M, 1) = 0.
\label{eq20} 
\end{equation}

\end{enumerate} 

Due to those properties, as one increases the time interval $\tau$
from $1$, the initial (truncated) power law distribution 
Eq. (\ref{eq17}) $R(>r,1) = S(>r)$ is not significantly affected 
for most of the $r$ range. For small $\tau$'s, the corrections to 
the power law are in fact limited only to the lower and upper 
cut-off regions and are analyzed below. 

\subsubsection{The low $r$ "dome"-like region}

The low $r$ region in $R(>r,\tau)$ is affected even for small
$\tau >1$ because there is a large probability that all of the 
$\tau$ steps are of the order $s_{mean}$ and lower. 
Consequently, the probability $R(>r,\tau)$ for values 
$r < \tau s_{mean} = \tau M/N$ is not given anymore 
by the probability $R(>r,1) = S(>r)$ of obtaining it through 
a single step, but rather by a combinatoric sum of probabilities 
of having $\tau$ (small, positive and negative) steps summing 
up to $r$. This is of course very similar to the way one estimates 
(through Poisson/Binomial expansion) the probability of a distance 
$r$ after a $\tau$ steps Gaussian walk. The consequence is a 
smoothening of the sharp end at $R(r=0,\tau)$. 
This concretizes in the appearance of a "dome"-like 
shape in the central region (around $r=0$) of the $R(r,\tau)$ 
probability density.

To estimate (the time dependence of) the extent of the "dome"-like 
region, we demand that the probability of achieving distances $r$ 
through $2$ or more steps is larger than the probability of 
achieving it through one step. Since $(\tau S(r/2))^2$ is the 
probability of having during $\tau$ steps at least $2$ steps of 
sizes at least $r/2$, and $\tau S(r)$ is the probability 
of having at least one of the $\tau$ steps of size at least $r$,
the condition describing the "dome"-like region is: 
$$
(\tau S(r/2))^2  > \tau S(r),
$$
which by substituting the power law Eq. (\ref{eq15}) for 
$S(>s)$ becomes: 
$$
\{\tau [r/(2s_{min})]^{-\alpha} \}^2 > \tau (r/s_{min})^{-\alpha}, 
$$
gives the central dome region: 
\begin{equation} 
r < 4 s_{min} \tau ^{1/\alpha} = 4 q M/N  \tau ^{1/\alpha}.
\label{eq21}
\end{equation} 
since $q$ is in (finance markets) practice not far from $1/4$ 
we denote 
\begin{equation} 
r_{dome} = s_{mean} \tau ^{1/\alpha}  = M/N \tau ^{1/\alpha}  
\label{eq22} 
\end{equation} 
and with this notation, the condition Eq. (\ref{eq21}) becomes 
\begin{equation} 
r < r_{dome}. 
\label{eq23} 
\end{equation} 
Thus, the power law Eq. (\ref{eq17}) remains unchanged in the range
\begin{equation}
r_{dome} < r < M,
\label{eq24}
\end{equation}
and the power law region disappears completely when the upper and 
lower limits of this interval coincide: 
$$
r_{dome}= M,
$$
i.e. (cf Eq. (\ref{eq22}) by the time that
$$
\tau ^{1/\alpha} M/N =  M 
$$
holds. This gives the maximal time for which one still has a 
power-like region:
\begin{equation}
\tau_{max}  = N^{\alpha}.
\label{eq25}
\end{equation}
Since the power law region is the crucial feature of the (truncated)
Levy distribution, $\tau_{max}$ is essentially the maximal time 
interval for which the returns (\ref{eq11}) still maintain a 
Levy-like probability distribution. Note that this value for
$\tau_{max}$ agrees with the early estimations of Ref. 
\cite{Mantegna94} based on the scaling $R(0,\tau)\sim \tau^{1/\alpha}$
of the probability distribution peak. 
 
\subsubsection{The extremely large $r> M$ region}

In the upper cut-off region $r>M$ the probability distribution 
function $R(>r,\tau)$ is affected by increasing $\tau$ from $1$ to 
larger values. Indeed, while values $r > M$ are completely disallowed 
for $\tau =1$ because of the truncation Eq. (\ref{eq20}), for 
$\tau >1$ one {\it can} have $r$'s in the range $M < r < \tau M$. 
The probability of obtaining such values of $r$ corresponds to the 
probability of selecting repeatedly for updating (by Eq. (\ref{eq5}))
the largest $w_i$'s. For instance, the probability of obtaining a value 
$r > K M $ (with $K$ a small integer) is roughly the probability 
of extracting out of the $\tau$ steps at least $2 K$ times steps of 
size at least $M/2$. For $K<< \tau$ this is basically the probability 
of at least one step of size at least $M/2$ multiplied by $\tau$ and 
risen at the power $2 K$: 
\begin{eqnarray}
R(>r=KM,\tau) &\sim& [\tau S(>M/2)]^{2K} \nonumber\\
&\sim& [\tau (M/(2 s_{min}))^{-\alpha}]^{2K}, \nonumber
\end{eqnarray}
which can also be written as
\begin{equation} 
R(>r,\tau) \sim \left [ \left ({N \over {(2q)^{\alpha}\tau}}
\right )^2 \right ]^{-r/M},
\label{eq26} 
\end{equation} 
i.e., $R(>r,\tau)$ decays exponentially with $r$. 
Values larger than $KM$ are still rigorously disallowed. 

To sum up the results for small and moderate intervals  
$\tau < \tau_{max}$: except for the central dome region 
$|r| < r_{dome}$ and the extremely large $|r|> M$ region, 
the probability distribution function $R(>r,\tau)$ is similar to 
the single step probability $R(>r,1)= S(>r)$ (Eq. (\ref{eq17})). 
This is intuitively explained by the fact that the probability to 
arrive after $\tau$ random steps at large (but less than $M$) 
$r$ values is dominated by the probability of having a single
step of order $r$.

The 3 regions above: central dome region $r<r_{dome}$, the Levy-like 
power law region, and the extremely large $r>M$ exponential region 
are the main features of the distribution $R(>r,\tau)$ for intervals 
$\tau < \tau_{max}$.

\subsection{Cross-over for $\tau > \tau_{max}$}
\label{sec2.4}

The time evolution of the $R(>r,\tau)$ shape depends on the 
fact that while the upper cut-off region (beyond which 
the power law fails) $r>M$ is fixed, the "dome" region expands 
with $\tau$ according to Eq. (\ref{eq22}). (Intuitively this 
is because, as one increases the number of time steps $\tau$, one 
can reach larger values of their sum $r(\tau)$ even if each of the 
individual steps is of order $\bar{w}$ or less). 
As seen above, this leads for intervals $\tau > \tau_{max}$ 
(Eq. (\ref{eq25})) to the disappearance of the intermediate 
Levy-like power law region in the $R(>r,\tau)$ distribution. 
After this time, the central dome will keep expanding 
on the expense of the $r>M$ cut-off region.

As it expands, the dome will assume a shape closer and closer to 
a Gaussian. This will be consistent with the central limit theorem 
as the involved elementary steps will be ultimately distributed on 
a finite support $s_{min}<s< M$ of quite limited extent compared 
with the $r \gg M$ range of values probed by the dome for very 
large times. 

Indeed, for time intervals $\tau \gg \tau_{max}$, the probability 
of many steps of size close to $M$ is not negligible. 
Thus the dynamics consists in a random walk of individual steps 
of size distributed within the finite support between $0$ and $M$. 
The distribution becomes a Gaussian whose expansion is dominated 
by the largest steps $M$. Since there will be roughly one step of 
size $M$ per $\tau_{max} = N^{\alpha}$ interval, the Gaussian 
width will expand as 
\begin{equation} 
\sigma(\tau) = M (\tau / \tau_{max})^{1/2} 
= M N^{-\alpha /2} \tau^{1/2}. 
\label{eq27} 
\end{equation}
For returns much larger than this, i.e., $r>\sigma (\tau)$, the
exponential regime Eq. (\ref{eq26}) will still survive.

The above results can be verified by the numerical simulations
shown below. Moreover, for time interval $\tau >\tau_{max}$ but
not too large, the power law distribution of returns with 
exponent well outside the stable Levy regime of $\alpha<2$ can 
be obtained in the simulations, with the exponential cut-off 
effect.

\begin{figure*}
\centerline{
\resizebox{0.7\textwidth}{!}{%
  \includegraphics[angle=-90]{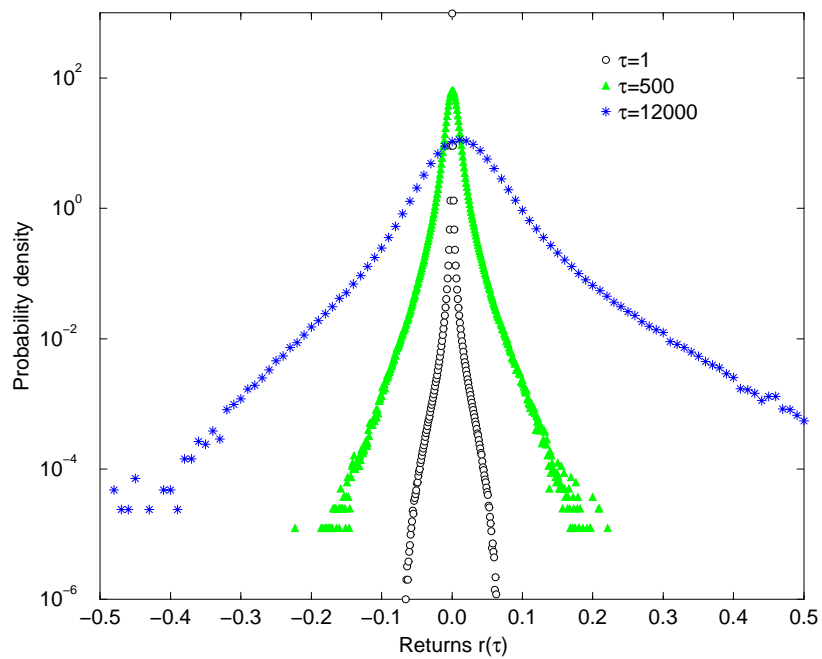}}}
\caption{Semi-plot of the probability density of returns
$r(\tau)$ (defined by Eq. (\ref{eq28})) for different
time intervals $\tau =1$, $500$, and $12000$ ($N=500$).}
\label{fig0}
\end{figure*}

\begin{figure*}
\centerline{
\resizebox{0.7\textwidth}{!}{%
  \includegraphics[angle=-90]{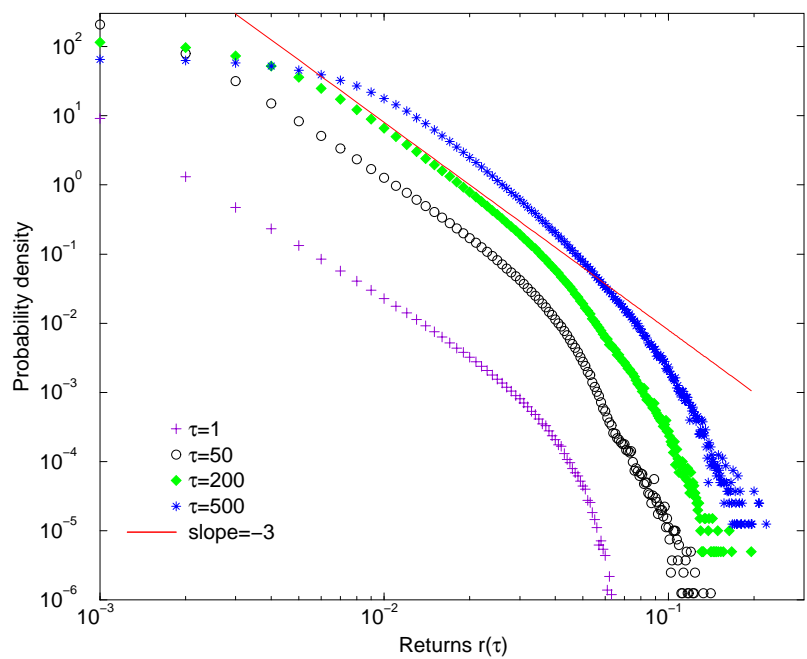}}}
\caption{The probability density of the positive returns $r(\tau)$
(defined by Eq. (\ref{eq28})) in log-log plot for time intervals 
$\tau =1$, $50$, $200$, and $500$ ($N=500$, calculated after 
$t=10^5 N$ updatings), with a straight line of slope $-3$ 
($\alpha=2$) for comparison.}
\label{fig1}
\end{figure*}

\section{Numerical simulations}

We have performed the computer simulations of autocatalytic
system described by Eqs. (\ref{eq5}) and (\ref{eq6}), where
the random factor $\lambda$ is set to uniformly distribute
in the range $0.9 < \lambda < 1.1$. In our simulations the
number of subsystems $N=500$, and the lower bound factor
$q=0.3$. Thus, according to relation (\ref{eq8}), the 
exponent $\alpha$ of the power law distribution of $w$
is about $1.4$, which has been verified by previous
simulations \cite{Malcai99}.

Here we numerically study the distribution of the fluctuations
or "returns"
\begin{equation}
r(\tau)=\ln W(t+\tau) - \ln W(t)
\label{eq28}
\end{equation}
for different time intervals $\tau$, to compare with the
above analytic results. Note that in Eq. (\ref{eq28}) we use
the logarithmic difference for the definition of return, as
in usual financial applications, which is approximately the
relative change Eq. (\ref{eq11}) if the change is small.
Note that due to the lower bound $q \bar{w}$ (Eq. (\ref{eq6})),
in this system $W(t)$ has an increasing trend, which makes
the return (\ref{eq28}) more possible to be positive,
especially for large $\tau$. Thus, in our simulation results
shown below, a maximum at positive finite $r$ is obtained
for large $\tau$ distribution.
In general, one could normalize $W(t)$ ($w_i(t)$) by a value
$\exp(\kappa t/N)$ with $\kappa$ constant, for detrending, and
then Eq. (\ref{eq28}) would change by a constant: $r(\tau) 
\rightarrow r(\tau)-\kappa \tau /N$, which does not influence 
the statistical properties.

With $N=500$ and $\alpha \sim 1.4$, we have 
$\tau_{max} = N^{\alpha} \sim 6000$ from Eq. (\ref{eq25}),
and the behaviors obtained for small and large $\tau$ are 
different, as shown in Fig. \ref{fig0} for $\tau =1$, $500$,
and $12000$. With the increase of time interval $\tau$, the
sharp peak of the distribution curve is smeared out into
a dome-like shape, similar to the empirical findings of
financial markets \cite{Gopikrishnan99}.

\subsection{$\tau < \tau_{max}$}

The numerical results of probability density $R(r,\tau)$ 
for time intervals $\tau=1$ (pluses), $50$ (circles), 
$200$ (diamonds) and
$500$ (stars) are presented in Fig. \ref{fig1}. The measurement
was performed after $t=10^5 N$ updatings, and averaged over
$100$ runs for $\tau =1$ and $800$ runs for $\tau \geq 50$.
The results for $\tau =1$ are just what we derive in Sec.
\ref{sec2.2}, i.e., a straight line in log-log plot with
sharp end, obeying Eq. (\ref{eq17}) with $\alpha$ about $1.5$, 
and a cut-off for large $r$.

For intervals larger than $1$, the 3 regions obtained
analytically in Sec. \ref{sec2.3} are clearly shown in
the log-log plots of Fig. \ref{fig1}. The first 
one is the central "dome"-like region for small $r$, with larger
extent for larger $\tau$, as predicted in Eq. (\ref{eq22}).
For small interval $\tau=50$, the derivative at small $r$ is
not close to zero, that is, the sharp end persists, while
it flattens into the dome for larger intervals of $200$
and $500$.
In the intermediate $r$ range, the power law behavior similar
to Eq. (\ref{eq17}) is presented. For small values of $\tau$
($\leq 200$ in Fig. \ref{fig1}), the exponent $\alpha$ is within 
the stable Levy regime, that is, $0<\alpha <2$, however, for 
larger $\tau$ (say $500$) one can obtain the exponent 
$\alpha >2$, which is similar to the phenomenon shown below
for $\tau > \tau_{max}$.

When the return $r$ is large, the deviation from the straight
line and the curvature in log-log plot can be observed (see
Fig. \ref{fig1}), which is just the cut-off effect described
above, that is, the exponential decay for far tail distribution
(similar to Eq. (\ref{eq26})).

\begin{figure*}
\centerline{
\resizebox{0.7\textwidth}{!}{%
  \includegraphics[angle=-90]{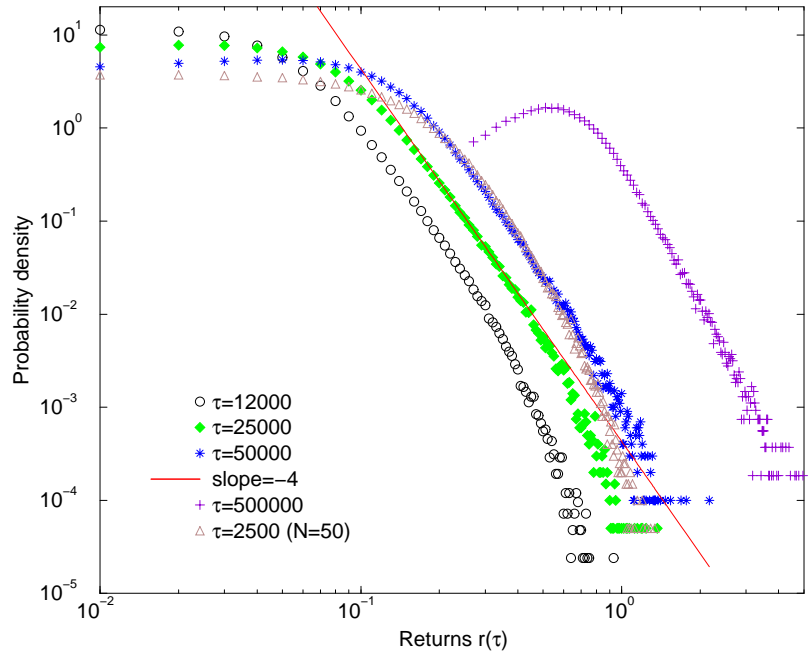}}}
\caption{The probability density of the positive returns $r(\tau)$
in log-log plot for large time 
intervals $\tau =12000$, $25000$, $50000$, and $500000$ of
$N=500$, as well as $\tau=2500$ of $N=50$
(calculated after $t=10^5 N$ updatings), with a straight line of 
slope $-4$ ($\alpha=3$) which fits the data for a few orders of 
magnitude. For $\tau =500000$, only part of the distribution is 
shown.}
\label{fig2}
\end{figure*}

\begin{figure*}
\centerline{
\resizebox{0.7\textwidth}{!}{%
  \includegraphics[angle=-90]{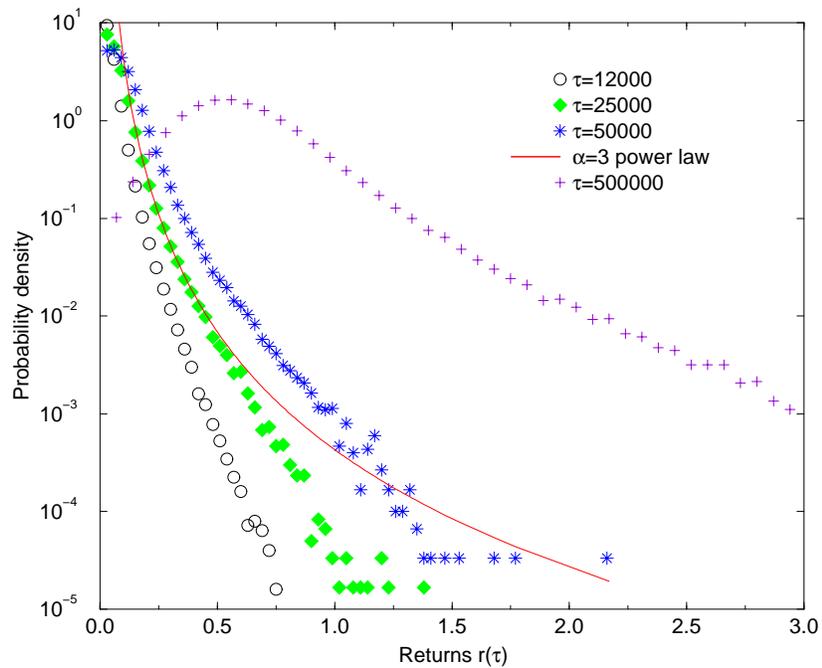}}}
\caption{Replot of Fig. \ref{fig2} on a semi-log scale, using
larger bin size.}
\label{fig3}
\end{figure*}

\begin{figure*}
\centerline{
\resizebox{0.7\textwidth}{!}{%
  \includegraphics[angle=-90]{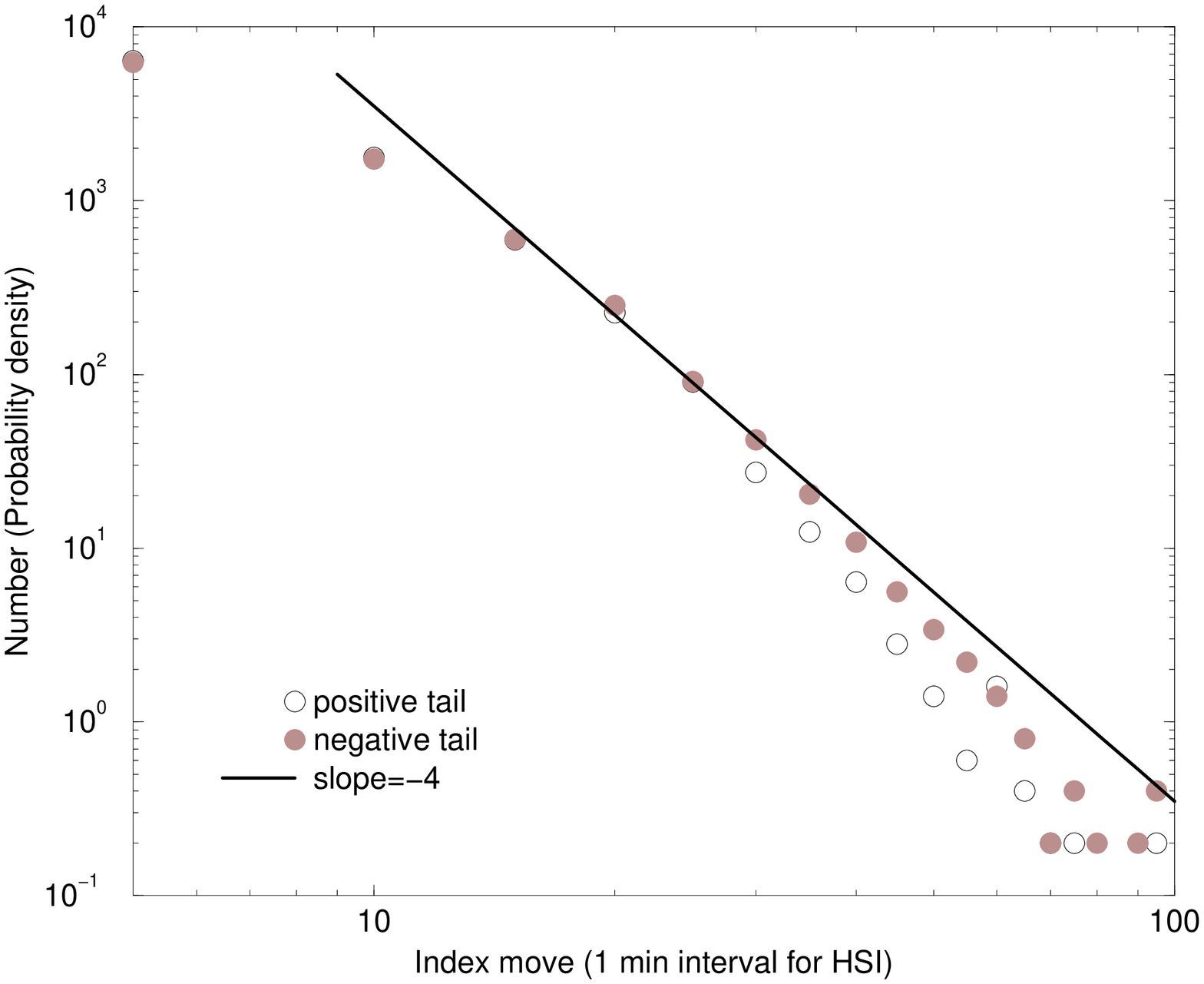}}}
\caption{Log-log plot of the probability density of 1 minute
index moves for the Hong Kong HSI (from Jan. 1994 to Dec. 1997),
with the skip of the first 20 minutes in daily opening 
\cite{Huang00}. A straight line with slope $-4$ is also shown.}
\label{fig4}
\end{figure*}

\subsection{$\tau > \tau_{max}$}

The crucial results of this paper are for large time interval
$\tau$, which corresponds to realistic time scale observed
in nature (e.g., financial markets). These results are the
consequences of the truncation in this autocatalytic system,
as discussed analytically above, and can be compared with 
those of the real market data.

We have calculated the return distribution for large $\tau$:
$12000$, $25000$, and $50000$ (averaged over $1000$ runs), 
as shown in Fig. \ref{fig2}. Besides the dome-like shape
for small $r$, what interests us is the power law region
for intermediate and large returns. In this region the 
exponent $\alpha$ is about $3$, well beyond the stable
Levy regime, but in agreement with the recent observations in
real stock market \cite{Lux96,Gopikrishnan99}.

However, for extremely large returns, i.e., the far tail of
the distribution, the exponential cut-off effect still 
remains, as shown in the bent of the log-log plots in Fig.
\ref{fig2}. To see more clearly, we replot Fig. \ref{fig2}
on a semi-log scale to Fig. \ref{fig3}, where the tail
of return distribution looks like a straight line,
indicating an exponential-type behavior.
This exponentially asymptotic decay was not
detected in the empirical findings of \cite{Lux96} and
\cite{Gopikrishnan99}, but agrees with the most recent
observation in Hong Kong stock market \cite{Huang00}
as shown in Sec. \ref{sec4} below.

To see whether this phenomenon of $\alpha=3$ power law
and exponential cut-off is related to finite size effect
or is intrinsic, we perform the simulations for $N=50$, 
and present the result of $\tau=2500$ in Fig. \ref{fig2}.
The range of power law behavior ($\alpha \sim 4.5$ for 
$N=50$) may be slightly shorter than that of the larger 
system $N=500$. Thus, the power law region before the
exponential cut-off is expected to extend for large system 
size ($N$), which has been verified by simulations for
$N=5000$.

For very large $\tau$, i.e., $\tau \gg \tau_{max}$, the
distribution of returns is to approach Gaussian based
on the central limit theorem (also expected in Sec. 
\ref{sec2.4}), which has been found in reality 
\cite{Gopikrishnan99}. In Fig. \ref{fig2} the result for
$\tau =500000$ is also shown with the Gaussian-like
behavior for not too large $r$ value, which exhibits as 
a parabola in semi-log plot (Fig. \ref{fig3}) \cite{note3}. 
The exponential-type decay is still found in Fig. \ref{fig3} 
for extremely large $r$ as expected.

\section{Discussion and conclusion}
\label{sec4}

In finance, the tradings of the various investors $w_i$ 
are performed independently. Therefore, the natural time
measure is not the number of operations but the number of
operations divided by the number of components $N$:
\begin{equation}
T=\tau /N.
\label{eq29}
\end{equation}

Thus, for the power law behavior of large $\tau$ ($500$,
$12000$, $25000$, and $50000$) shown in Figs. \ref{fig1} 
and \ref{fig2} with exponent well outside the Levy regime,
the corresponding $T$ values are $1$, $24$, $50$, and $100$
(for $N=500$ in our simulations).
Although in real market the time interval between transactions
of stocks is irregular, one could estimate from the market
transaction data \cite{Plerou99} that the unit scale
$T=1$ here approximately corresponds to several minutes of
real time.

It is interesting to compare the results of this autocatalytic 
system with actual measurements of the stock markets. The
power law behavior with exponent $\alpha$ about $3$ was observed
in recent empirical studies on S\&P 500 \cite{Gopikrishnan99}
and German DAX \cite{Lux96}, consistent with the results 
shown in Fig. \ref{fig2} for intermediate and large returns,
but not the exponential cut-off. However, very recently it
has been found from the Hang Seng Index (HSI) of Hong Kong 
\cite{Huang00} that the index fluctuations for the first
few minutes of daily opening behave very differently from
those of the other times, due to much higher influences of
exogenous factors at the opening. As shown in Fig. \ref{fig4},
if we skip the data in the first $20$ minutes of each trading
day, the distribution for 1 minute time interval index move
(defined as ${\rm index}(t+\tau)-{\rm index}(t)$, with the 
statistical properties very similar to that of Eq. (\ref{eq28}) 
for the high-frequency regime, e.g., interval $\tau =1$ min)
exhibits the phenomenon of exponential-type decay \cite{Tang00} 
after the transient $\alpha=3$ power law region, in agreement 
with our simulation results of Figs. \ref{fig3} and \ref{fig1}
($\tau=500$). Moreover, real markets show the tendency of a 
crossover towards a Gaussian for long enough times 
\cite{Gopikrishnan99}, which has also been found in our system
(see Figs. \ref{fig2} and \ref{fig3} for very large interval
$\tau=500000$ and intermediate $r$).

In order to account for the experimentally observed volatility
correlations, one may follow Ref. \cite{Solomon98} and feed back 
the absolute market returns (Eq. (\ref{eq28})) into the individual 
gain factor (\ref{eq5}).
A possible form is using:
$$
\lambda (t) = \exp[\langle r \rangle 
+  \eta ( \langle r^2 \rangle  -  \langle r \rangle^2)^{1/2} ]
$$
where the angle brakets indicate averages over the
last $\tau$ steps and $\eta$ is a Gaussian random number of zero 
mean and unit variance.
The numerical results for these more realistic simulations
will be presented elsewhere.

In summary, we have shown that the simple random multiplicative 
model of Ref. \cite{Solomon96}, with a lower cut-off, gives many 
of the properties found in reality and in more complicated models, 
like e.g. the percolation model \cite{Cont00}: power law with 
effective exponent $\alpha$ near 3, rounding of the singularity at 
zero returns, crossover to Gaussian for long times. We still have to 
work on implementing volatility clustering, multifractality, lack 
of up-down and time reversal symmetry, and correlation between 
traded volume and volatility. Only the original random walk model 
of Bachelier seems to us simpler than the present model but of the 
above properties it has only a Gaussian distribution. 

\begin{acknowledgement}
We thank Dietrich Stauffer
for very helpful discussions and comments. Z.F.H. acknowledges the
financial support of SFB 341 and computer support of German-Israeli 
Foundation. One of us (S.S.) would like to acknowledge
many discussions over the last years and common
research on the subject with O. Biham and O. Malcai.
\end{acknowledgement}


\begin{thebibliography}{}
\bibitem{comment1} In order not to complicate the notation we will
not write $s_{i_k}$ in Eq. (\ref{eq12}), but one should keep in
mind that the $w_i$ selected for updating changes after each time
step.
\bibitem{note3} The phenomenon that the parabola is shifted to the right 
is related to the fact that we used a large value for the volatility
to reach faster the convergence of the numerical runs.
In real financial systems this shift is unobservable for
intraday fluctuations.
\end{thebibliography}
\end{document}